\documentclass[cameraready]{Interspeech}

\title{ArtNet: A JEPA-Like Articulatory Predictive Framework for Robust Zero-Shot Phoneme Recognition}

\author[affiliation={1}, orcid=0000-0002-8597-2882, equalcontribution]{Zeqian}{Hu}
\author[affiliation={2}, equalcontribution]{Fuliang}{Weng}
\author[affiliation={1}, orcid=0009-0000-8640-5479]{Shu}{Shang}
\author[affiliation={1}, correspondingauthor]{Yaqian}{Zhou}

\address{
    $^1$ Fudan University, China \\
    $^2$ Pedawise, Shanghai, China
}

\email{zqhu24@m.fudan.edu.cn, fuliang.weng@pedawise.cn, zhouyaqian@fudan.edu.cn}


\usepackage{float}
\usepackage{comment}
\usepackage{amsmath}
\usepackage{graphicx}  
\usepackage{booktabs}  
\usepackage{multirow}  
\usepackage[capitalize]{cleveref}

\graphicspath{{figure/}}

\begin{document}

\maketitle

\begin{abstract}

Zero-shot cross-lingual phoneme recognition is often hindered by the fragility of direct acoustic-to-symbol mapping, which is susceptible to language-specific variations. Echoing joint-embedding predictive architecture (JEPA) work in vision, we propose ArtNet, a framework that explores a structured feature prediction task based on articulatory features to enhance acoustic robustness. Specifically, ArtNet integrates an articulatory predictor—designed to extract universal articulatory representations from self-supervised learning (SSL) features—with a variational information bottleneck (VIB) to suppress language-specific variations. Experiments on seven unseen languages demonstrate that ArtNet, particularly when synergized with the proposed vector-space inventory alignment (VSIA) strategy, significantly outperforms competitive baselines, achieving a 20.56\% relative reduction in phoneme error rate (PER) and 7.01\% in phoneme feature error rate (PFER).

\end{abstract}

\section{Introduction}

Modern end-to-end (E2E) automatic speech recognition (ASR) systems have achieved remarkable success in resource-rich languages such as English and Mandarin. However, the vast majority of the world’s languages lack sufficient paired audio-text data, posing a significant challenge for E2E models to achieve comparable precision.

In multilingual and cross-lingual scenarios, a more data-efficient paradigm involves developing a robust phoneme recognizer as an intermediate stage \cite{yusuyin2025whistle}. The prevailing architecture for training such phoneme recognizers typically follows an encoder-decoder framework. Under this paradigm, the encoder is first pre-trained through self-supervised learning like Wav2Vec2 \cite{baevski2020wav2vec} and HuBERT \cite{hsu2021hubert}, followed by a joint fine-tuning of both the encoder and decoder on a limited amount of paired audio-text data. Such a framework is hereafter referred to as an SSL-based phoneme recognizer.

\begin{figure}[t]
  \centering
  \includegraphics[width=\linewidth]{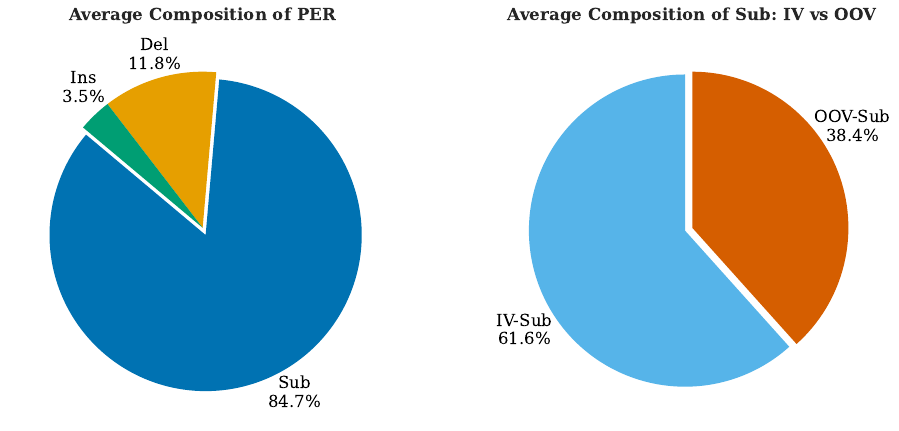}
  \caption{Detailed error analysis of the phoneme recognizer in zero-shot scenario. The left chart illustrates the macro-average decomposition of the PER into Substitution (Sub), Deletion (Del), and Insertion (Ins) error rate. The right chart provides a breakdown of substitution errors, distinguishing between in-vocabulary (IV) and out-of-vocabulary (OOV) phonemes across all unseen languages.}
  \label{fig:error_decomposition_summary}
\end{figure}

However, empirical analysis suggests of our phoneme recognizer trained on English that the primary bottleneck in cross-lingual generalization stems not only from the presence of unseen phonemes but, more significantly, from the brittleness of previously learned phonetic representations. As illustrated in \cref{fig:error_decomposition_summary}(left), the macro-average decomposition of the PER across seven unseen languages indicates that the substitution error is the predominant component, accounting for 84.7\% of the total errors. This confirms that the challenge lies in the categorical discriminability of the phonetic frontend.

Crucially, a deeper examination of these substitutions in \cref{fig:error_decomposition_summary}(right) reveals a counter-intuitive finding: while one might expect out-of-vocabulary (OOV) phonemes to be the primary source of error in zero-shot scenarios, 61.6\% of substitutions are actually attributed to in-vocabulary (IV) phonemes.

These observations suggest that the performance degradation of the cross-lingual recognition models is largely driven by the lack of discrimination at the shared phoneme representation level. Although existing work \cite{gao2021zero} has attempted to incorporate auxiliary linguistic knowledge to mitigate this, they often overlook the need to improve the intrinsic robustness of the phoneme recognizer itself. In line with the philosophy of JEPA in computer vision \cite{assran2023self} and its recent extensions to the audio domain \cite{ioannides2025jepa, tuncay2025audio}, we posit that this vulnerability stems from the prevailing training paradigm that attempts to map raw acoustic signals directly onto discrete symbols—a process inherently susceptible to non-essential acoustic fluctuations and at the surface level. Therefore, rather than focusing on OOV mapping, a more critical priority is to shift toward a predictive modeling framework that extracts language-invariant features within a structured representation space, thereby strengthening the discriminative robustness of IV phonemes across diverse linguistic environments.

The articulatory features, which describe the physical movements of speech organs during sound production, offer a potential solution to this generalization gap. This perspective aligns with the foundational theory of \textit{The Sound Pattern of English} \cite{chomsky1968sound}, which posits that phonemes are not atomic symbols but bundles of universal distinctive features grounded in human physiology. Unlike acoustic signals, articulatory features are fundamentally shared between languages. Existing frameworks have utilized these features as a universal bridge by decomposing phonemes into shared attribute spaces or employing phonological-vector based phone embeddings \cite{li2020towards,zhu2021multilingual,lee2023embedding,yen2024boosting}. Although promising, these models often fail to disentangle linguistic representations from language-specific acoustic characteristics and prosodic patterns. Consequently, they risk overfitting to source linguistic environments, which limits their robustness when encountering truly unseen languages with significantly different acoustic variations.

To address the limitations, we propose a novel framework to solve these problems. The primary contributions of this research are summarized as follows:

\begin{itemize}
\item \textbf{ArtNet: A JEPA-like articulatory framework:} Moving beyond the conventional paradigm of direct acoustic-to-symbol mapping, we reformulate cross-lingual phoneme recognition as a non-generative predictive task. By utilizing a specialized \textbf{articulatory predictor} to map SSL features into a structured, physical articulatory space, our model learns representations that are inherently more stable across linguistic boundaries than raw acoustic embeddings.
  
\item \textbf{Mitigation of the substitution error bottleneck:} We demonstrate that \textbf{ArtNet} successfully mitigates the predominant substitution errors by filtering out language-specific variations. This ensures that the extracted articulatory cues remain robust and discriminative even when encountering significantly different acoustic context.
  
\item \textbf{Novel inference strategy and superior performance:} To maximize the utility of the learned articulatory space, we introduce \textbf{vector-space inventory alignment (VSIA)}, as a zero-shot inference strategy. Driven by this methodological design, our approach achieves an average relative reduction of 20.56\% in PER and 7.01\% in PFER over a competitive SSL-based baseline across seven unseen languages.
\end{itemize}

\section{Methodology}

\subsection{Construction of Structured Articulatory Targets}

To enable the explicit modeling of the articulatory space, we construct a structured target space that provides supervision without the need for manual phonetic annotations. Within this framework, the SSL backbone functions as a context encoder, which is first optimized via a connectionist temporal classification (CTC) objective to establish a foundational phoneme recognizer. To unify the label space, orthographic transcripts are transduced into international phonetic alphabet (IPA) sequences using the Epitran G2P library \cite{mortensen2018epitran}. 

Once initialized, the phoneme recognizer serves as a pseudo-label generator that provides stable, frame-level phonetic alignments necessary for guiding the subsequent predictive learning process. Let $\mathbf{H} = (\mathbf{h}_1, \dots, \mathbf{h}_T)$ denote the sequence of frame-level hidden representations extracted from the context encoder. For each frame, the pseudo-label generator produces a phoneme $\pi_t \in \Phi$ via greedy decoding, where $\Phi$ represents the universal phoneme inventory. 

To facilitate learning in a structured target space, we utilize the Panphon database \cite{mortensen2016panphon} to decompose each IPA symbol into a 24-dimensional vector of articulatory attributes, originally represented by the trinary values $\{+, -, 0\}$. For network compatibility, we transform these categorical labels into a numerical encoding scheme where $\{+\}$ is mapped to $1$, $\{-\}$ to $-1$, and $\{0\}$ remains $0$. This process yields a static mapping matrix $\mathbf{M} \in \{-1, 0, 1\}^{|\Phi| \times 24}$. Consequently, the ground-truth articulatory vector $\mathbf{v}_t$ for each time step is derived via a mapping function $\mathcal{M}(\cdot)$ that performs a row-lookup on $\mathbf{M}$ according to the generated pseudo-label $\pi_t$:

\begin{equation}
  \mathbf{v}_t = \mathcal{M}(\pi_t)
  \label{equation:eq1}
\end{equation}

By mapping the discrete pseudo-labels into this continuous, structured target space, we provide language-agnostic supervision signals. This predictive paradigm forces the model to disentangle linguistic content from the language-specific variations inherent in the encoder representations, establishing a more universal foundation for cross-lingual transfer.

\subsection{ArtNet}

To further improve the robustness of feature disentanglement, we propose ArtNet, a dedicated neural framework for structured articulatory modeling. ArtNet is composed of an articulatory predictor (AP) equipped with a variational information bottleneck (VIB). The AP is designed as a lightweight module to predict articulatory vectors from a robust latent space. Instead of feeding the hidden state $\mathbf{h}_t$ directly to the AP, we maps it to a stochastic latent encoding $\mathbf{z}_t$ to suppress noise.

We assume the latent distribution $p(\mathbf{z}_t|\mathbf{h}_t)$ follows a multivariate gaussian distribution $\mathcal{N}(\boldsymbol{\mu}_t, \boldsymbol{\sigma}_t^2)$. The parameters are predicted by the VIB encoder, where $f_{\mu}(\cdot)$ and $f_{\sigma}(\cdot)$ are implemented as fully connected layers:

\begin{equation}
  \boldsymbol{\mu}_t = f_{\mu}(\mathbf{h}_t),
  \quad
  \boldsymbol{\sigma}_t = f_{\sigma}(\mathbf{h}_t)
  \label{equation:eq2}
\end{equation}

Using the reparameterization trick, we sample $\mathbf{z}_t$ as:

\begin{equation}
  \mathbf{z}_t = \boldsymbol{\mu}_t + \boldsymbol{\sigma}_t \odot \boldsymbol{\epsilon},
  \quad
  \boldsymbol{\epsilon} \sim \mathcal{N}(\mathbf{0}, \mathbf{I})
  \label{equation:eq3}
\end{equation}

The AP then predicts the articulatory vector $\hat{\mathbf{v}}_t$ from this latent variable:

\begin{equation}
  \hat{\mathbf{v}}_t = \text{AP}(\mathbf{z}_t)
  \label{equation:eq4}
\end{equation}

The loss of reconstruction $\mathcal{L}_{\mathrm{AP}}$ measures the discrepancy between the predicted articulatory vector $\hat{\mathbf{v}}_t$ and the ground truth $\mathbf{v}_t$. We employ the mean squared error (MSE) as the distance metric:

\begin{equation}
  \mathcal{L}_{\mathrm{AP}} = \frac{1}{T} \sum_{t=1}^{T} \parallel \mathbf{v}_t - \hat{\mathbf{v}}_t \parallel_2^2
  \label{equation:eq_artnet}
\end{equation}

Information bottleneck regularization is defined using the Kullback-Leibler (KL) divergence. Since the prior $r(\mathbf{z})$ is a standard normal distribution $\mathcal{N}(\mathbf{0}, \mathbf{I})$, the divergence can be computed analytically:

\begin{equation}
\begin{aligned}
  \mathcal{L}_{\mathrm{VIB}} &= \frac{1}{T} \sum_{t=1}^{T} D_{\mathrm{KL}}(p(\mathbf{z}_t|\mathbf{h}_t) \parallel r(\mathbf{z}))
\end{aligned}
  \label{equation:eq_vib_loss}
\end{equation}

Finally, the total objective function integrates the reconstruction loss and the regularization via a hyperparameter $\beta$:

\begin{equation}
  \mathcal{L} = \mathcal{L}_{\mathrm{AP}} + \beta \mathcal{L}_{\mathrm{VIB}}
  \label{equation:eq_total_loss}
\end{equation}

\subsection{Zero-Shot Inference Strategy}

During the inference phase, we adopt a hybrid decoding strategy designed to handle the constraints of zero-shot scenarios. The CTC decoder first determines the topology of the sequence. For any time step $t$ where the CTC prediction $y^{ctc}_t$ is $\langle \text{blank} \rangle$, the frame is considered devoid of articulatory content and is explicitly skipped.

\subsubsection{Pooling Inference}
In zero-shot cross-lingual settings, the CTC model often exhibits limited capability in precisely determining phoneme boundaries, rendering frame-by-frame inference susceptible to noise and instability. To mitigate this, we introduce a pooling inference strategy. Instead of predicting articulatory features for every single frame, we group consecutive frames that are predicted as the same non-blank class into a segment. For each segment $S$, we compute the mean of the corresponding encoder output vectors $\mathbf{h}_t$:

\begin{equation}
  \bar{\mathbf{h}}_S = \frac{1}{|S|} \sum_{t \in S} \mathbf{h}_t
  \label{equation:eq_hs}
\end{equation}

This aggregated representation $\bar{\mathbf{h}}_S$ is then fed into AP to predict a robust articulatory vector $\hat{\mathbf{v}}_S$. Note that strictly speaking, $\hat{\mathbf{v}}_S$ resides in a continuous physical space and does not yet correspond to a discrete symbol.

\subsubsection{Vector-Space Inventory Alignment (VSIA)}
\label{subsubsec:vsia}

To overcome the rigid phoneme-level inventory constraints imposed by the \textit{tr2tgt} strategy \cite{xu2022simple}, we introduce VSIA to adapt continuous predictions directly to the target language. While the original strategy forces a many-to-one discrete mapping by minimizing the Hamming edit distance over binary articulatory attributes, our approach explicitly leverages the rich, continuous geometric structure learned by AP.

Specifically, we utilize the underlying structure of the target language's phoneme inventory, denoted as $\Phi_{tgt}$, to constrain the search space. Instead of discretizing the predicted vector $\hat{\mathbf{v}}_S$ immediately, we perform a nearest-neighbor search based on angular proximity in the continuous articulatory space. The final predicted phoneme $\hat{\phi}$ is identified as the candidate in $\Phi_{tgt}$ that maximizes the cosine similarity with the prediction:

\begin{equation}
    \hat{\phi} = \operatorname*{argmax}_{\phi \in \Phi_{tgt}} \left( \frac{\hat{\mathbf{v}}_S \cdot \mathcal{M}(\phi)}{\lVert\hat{\mathbf{v}}_S\rVert \lVert\mathcal{M}(\phi)\rVert} \right)
\label{eq:vsia_inference}
\end{equation}

Through soft alignment, VSIA mitigates phonological mismatches and ensures the output adheres to the target language's valid phonetic definitions.

\section{Experiment}

\subsection{Datasets}

To evaluate the cross-lingual generalization capability of our proposed method, we define distinct source and target domains.

\begin{itemize}
  \item \textbf{Training Data (Source)} We utilize the LibriSpeech \cite{panayotov2015librispeech} train-clean-100 dataset as our primary training corpus. This subset consists of approximately 100 hours of clean English speech.
  \item \textbf{Test Data (Target)} For zero-shot evaluation, we utilize the standard test sets of seven non-English languages (German, Dutch, French, Spanish, Italian, Portuguese, and Polish) from the Multilingual LibriSpeech (MLS) dataset\cite{pratap2020mls}.
  \item \textbf{Data Preprocessing} To unify the label space across diverse languages, we employ Epitran to convert the orthographic transcriptions of all datasets (both LibriSpeech and MLS) into the IPA symbols.
\end{itemize}

\subsection{Model Training and Implementation Details}

In this work, we employ mHuBERT-147\footnote{\url{https://huggingface.co/utter-project/mHuBERT-147}} \cite{zanon2024mhubert} as the SSL backbone for speech representation extraction. It is a 95M parameter model consisting of 12 Transformer encoder layers, with an embedding dimension of 768 and 12 attention heads. The model was pre-trained on an extensive multilingual corpus encompassing 147 different languages. 

During the phoneme recognizer initialization phase, we utilize low-rank adaptation (LoRA) to efficiently optimize the context encoder and CTC head. In the subsequent articulatory feature prediction phase, the phoneme recognizer's parameters are frozen, and training focuses exclusively on the AP and VIB modules. To investigate the role of temporal context in feature extraction, we evaluate ArtNet across three architectural variants: a Multi-Layer Perceptron (MLP), a Time Delay Neural Network (TDNN), and a Long Short-Term Memory (LSTM) network. These architectures correspond to distinct levels of temporal modeling—ranging from the context-independent MLP and the locally constrained TDNN to the LSTM, which captures long-range sequential dependencies. This comparison allows us to isolate the optimal level of temporal dynamics required for robust articulatory modeling. For the VIB configuration, the dimension of the latent variable $\mathbf{z}$ is set to $128$, with the trade-off parameter $\beta$ set to $0.001$ to balance reconstruction loss and KL-divergence regularization.

Both stages are optimized using the Adam optimizer \cite{kingma2014adam}. The learning rate is linearly warmed up for the first 10\% training steps  and peaks at 1e-3. To ensure a strict zero-shot evaluation, the entire framework is trained exclusively on the English source data.

\subsection{Evaluation Metrics}

Standard PER doesn't consider the phonemic differences
in terms of phonological similarity. To directly evaluate the quality of the predicted physical articulatory attributes, we use the PFER metric as described in \cite{taguchi2023universal}. PFER measures the distance between the predicted feature vectors and the ground-truth articulatory features, providing a more granular assessment of the model's phonetic disentanglement ability.

\begin{table*}[th]
    \centering
    \caption{Zero-shot cross-lingual performance comparison (in \%) measured by PER and PFER. ``tr2tgt'' denotes the phoneme mapping strategy, and ``ArtNet+VSIA'' represents the proposed method. The best results are highlighted in \textbf{bold}. $\downarrow$ indicates lower is better.}
    \label{tab:performance_comparison}
    \begin{tabular}{lcccccccc}
        \toprule
         & Dutch & French & German & Italian & Polish & Portuguese & Spanish & Avg \\
        \midrule
        \multicolumn{9}{l}{\textbf{PER}$\downarrow$} \\
        Baseline & 59.67 & 59.33 & 52.63 & 54.43 & 55.08 & 61.38 & 58.76 & 57.33 \\
        ArtNet & 58.64 & 56.84 & 51.63 & 51.73 & 50.24 & 59.94 & 55.57 & 54.94 \\
        Baseline+tr2tgt & 56.12 & 56.54 & 50.48 & 46.07 & 41.01 & 58.02 & 34.27 & 48.93 \\
        ArtNet+VSIA & \textbf{55.40} & \textbf{53.75} & \textbf{50.04} & \textbf{39.96} & \textbf{35.18} & \textbf{53.93} & \textbf{30.50} & \textbf{45.54} \\
        \midrule
        \multicolumn{9}{l}{\textbf{PFER}$\downarrow$} \\
        Baseline & 15.54 & 21.90 & 12.66 & 10.16 & 10.64 & 14.16 & 10.79 & 13.69 \\
        ArtNet & \textbf{15.33} & 21.76 & \textbf{12.49} & 9.81 & 10.21 & 13.89 & 10.33 & 13.40 \\
        Baseline+tr2tgt & 15.84 & 20.96 & 12.78 & 10.23 & 10.58 & 14.51 & 9.41 & 13.47 \\
        ArtNet+VSIA & 16.08 & \textbf{20.82} & 12.71 & \textbf{8.48} & \textbf{9.49} & \textbf{13.63} & \textbf{7.92} & \textbf{12.73} \\
        \bottomrule
    \end{tabular}
\end{table*}

\section{Results and Analysis}

\subsection{Zero-Shot Cross-Lingual Performance}

To evaluate the effectiveness of our proposed method in zero-shot scenarios, we conducted experiments on seven target languages using a model trained solely on English. The performance is measured using PER and PFER.

The main results, utilizing the TDNN-based ArtNet synergized with the VSIA strategy, are presented in Table \ref{tab:performance_comparison}. Compared to the standard SSL-based baseline, this combined approach demonstrates consistent improvements across the majority of target languages and metrics. Specifically, the integration of ArtNet and VSIA yielded an average relative reduction of 20.56\% in PER compared to the baseline. Notably, substantial gains are achieved in Romance languages such as Spanish, where the PER decreased by approximately 28.26\% absolute points. The overall reduction in PFER (averaging 7.01\% relative improvement) further confirms that our proposed framework, by aligning robust articulatory predictions with the target inventory, successfully captures more accurate articulatory features.

\subsection{Analysis of ArtNet}

It is important to note that both tr2tgt and VSIA strategy leverage external linguistic knowledge to align source and target phoneme spaces. To decouple this external benefit and rigorously evaluate the intrinsic robustness of the articulatory features learned by ArtNet, the following analyses focus on the model outputs before the application of zero-shot inference strategy.

\subsubsection{Generalization Analysis: Insights from Substitution Errors}

To assess adaptability, we analyze substitution errors (Table \ref{tab:sub_analysis}). Beyond the anticipated 5.67\% error reduction for in-vocabulary (IV) tokens, ArtNet surprisingly reduced out-of-vocabulary (OOV) errors by 1.62\%. This dual improvement indicates that rather than overfitting to training patterns, explicitly modeling articulatory features allows ArtNet to capture universal phonetic properties that generalize effectively to unseen tokens.

\begin{table}[th]
    \centering
    \caption{Analysis of the number of substitution error tokens.}
    \label{tab:sub_analysis}
    \begin{tabular}{lccc}
        \toprule
        Method & Total & IV & OOV \\
        \midrule
        Baseline & 1,078,964 & 665,073 & 413,891 \\
        ArtNet & 1,034,535 & 627,342 & 407,193 \\
        \midrule
        Relative Reduction & 4.12\% & 5.67\% & 1.62\% \\
        \bottomrule
    \end{tabular}
\end{table}

\subsubsection{Ablation Study: Impact of ArtNet Backbone Architecture}

To determine the optimal architecture for extracting articulatory features, we evaluate three ArtNet backbone variants: a 2-layer MLP (no context), a 2-layer TDNN (local context), and a 2-layer LSTM (global context). We compare these against the baseline to quantify the gain derived from articulatory modeling. The average PER across all seven target languages is summarized in Table \ref{table:backbone_ablation}.

As shown in the results, all ArtNet variants outperform the baseline. The experimental results, with TDNN ($54.94\%$) yielding the lowest error rate, followed by MLP ($55.33\%$) and LSTM ($56.51\%$), provide critical insights into cross-lingual modeling. 

Specifically, while LSTM is capable of modeling long-range dependencies, its higher error rate suggests that the global context from the source language contains language-specific phonotactic and prosodic biases. These biases likely act as high-level variations during zero-shot transfer to a target language with different rhythmic structures. In contrast, TDNN strikes an ideal balance by utilizing a constrained receptive field; it captures universal, local phonetic transitions while filtering out detrimental global biases. Furthermore, the fact that even the frame-level MLP outperforms the baseline underscores the inherent robustness of the articulatory space as a language-agnostic anchor, which proves more effective for cross-lingual mapping than raw acoustic features alone.

\begin{table}[th]
  \caption{Average PER (\%) of different ArtNet backbone architectures across seven target languages.}
  \centering
  \begin{tabular}{lcc}
    \toprule
    \textbf{Model / Backbone} & \textbf{Context Type} & \textbf{Avg. PER $\downarrow$} \\
    \midrule
    Baseline & - & 57.33 \\
    \midrule
    ArtNet (LSTM) & Global & 56.51 \\
    ArtNet (MLP)     & None   & 55.33 \\
    \textbf{ArtNet (TDNN)} & \textbf{Local} & \textbf{54.94} \\
    \bottomrule
  \end{tabular}
  \label{table:backbone_ablation}
\end{table}

\section{Conclusion}

In this work, we addressed the generalization challenge in zero-shot cross-lingual ASR by reformulating articulatory modeling as a JEPA-like structured prediction task. We proposed ArtNet, a framework that integrates an articulatory predictor with a VIB to project SSL representations into a universal, physical target space. This architecture proved critical in filtering out language-specific variations and implicitly handling structural redundancy, thereby significantly reducing substitution errors. Our comprehensive evaluations show that ArtNet, particularly when synergized with the proposed VSIA strategy, yields superior performance in terms of PER and PFER. Overall, our findings suggest that shifting towards non-generative, predictive modeling of articulatory features offers a promising direction for enhancing the intrinsic robustness of next-generation multilingual speech systems.

\section{Generative AI Use Disclosure}

During the preparation of this manuscript, the authors used Gemini solely for editing and polishing the English language. The tool was not used for producing any significant part of the manuscript.

\bibliographystyle{IEEEtran}
\bibliography{mybib}

\end{document}